# Spurious-Free Lithium Niobate Bulk Acoustic Resonator for Piezoelectric Power Conversion

Kristi Nguyen[†], Eric Stolt[‡], Weston Braun[‡], Vakhtang Chulukhadze[†], Jeronimo Segovia-Fernandez[§],
Sombuddha Chakraborty[§], Juan Rivas-Davila[‡], and Ruochen Lu[†]

[†]Department of Electrical and Computer Engineering, The University of Texas at Austin, Austin, US
[‡]Department of Electrical Engineering, Stanford University, Stanford, US
[§]Kilby Labs, Texas Instruments, Santa Clara, US
kristi.nguyen@utexas.edu

*Summary*—Recently, piezoelectric power conversion has shown great benefits from replacing the bulky and lossy magnetic inductor in a traditional power converter with a piezoelectric resonator due to its compact size and low loss. However, the converter performance is ultimately limited by existing resonator designs, specifically by moderate quality factor ($Q$), moderate electromechanical coupling ($k_t^2$), and spurious modes near resonance. This work reports a spurious-free lithium niobate (LiNbO$_3$) thickness-extensional mode bulk acoustic resonator design, demonstrating $Q$ of 4000 and $k_t^2$ of 30% with a fractional suppressed region of 62%. We first propose a novel grounded ring structure for spurious-free resonator design, then validate its performance experimentally. Upon further work, this design could be extended to applications requiring spurious suppression, such as filters, tunable oscillators, transformers, etc.

## I. Introduction

Due to their shorter acoustic wavelength and lower loss, acoustic devices have replaced their radio-frequency (RF) counterparts with commercial success in applications such as front-end filters and oscillators [1]–[9]. More recently, piezoelectric power conversion has emerged as yet another application, where inductors are replaced with acoustic resonators in power converters to reduce form factor and improve performance.

Piezoelectric power converter circuits are modeled as a resonator connected to various switch configurations ($S_1$, $S_2$, $S_3$, $S_4$) and direct current (DC) voltage sources ($V_{in}$, $V_{out}$) [Fig. 1 (a)]. The resonator is modeled with an equivalent electrical circuit called the Butterworth-Van Dyke (BVD) circuit that consists of a series motional inductor, resistor, and capacitor ($L_m$, $R_m$, $C_m$) connected in parallel with a static capacitance ($C_0$).

The converter's operation range is restricted by the resonator's inductive behavior, i.e., between series and parallel resonances. Although the converter is excited with DC voltages, zero-voltage switching sequences are leveraged to induce a motional current within the resonator at the operating frequency [Fig 1 (b)]. By tuning the switches' timings, the operating frequency can be varied, which ultimately determines the converter's output power for a given voltage conversion ratio. These switching sequences comprise *connected*, *open*, and *zero* stages that soft-charge the resonator's static

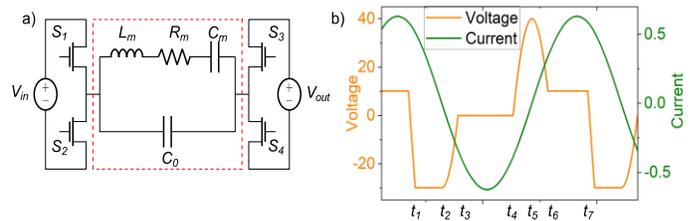

Fig. 1 (a) Circuit schematic of piezoelectric resonator, modeled by BVD circuit, integrated into power converter. (b) Idealized voltage and current waveforms through resonator for 40 V input, 30 V output [32].

TABLE I SOA OF PIEZOELECTRIC POWER CONVERTER RESONATORS

| Reference | $f_s$ (MHz) | $k_t^2$ | $Q$ | FoM | Spurious Supp. Region (MHz) | Fractional Supp. Region |
|---|---|---|---|---|---|---|
| PZT-radial [34] | 0.48 | 19% | 1030 | 196 | 0.015 | 42.9% |
| LN-TS [35] | 3.55 | 53% | N/A | N/A | N/A | N/A |
| LN-TS [36] | 5.94 | 45% | 3500 | 1575 | 0.37 | 34.9% |
| LN-TE [31] | 6.28 | 25.5% | 3700 | 944 | N/A | N/A |
| LN-TE [32] | 6.82 | 29% | 4178 | 1212 | 0.027 | 3.38% |
| **LN-TE [This work]** | **10.14** | **30%** | **4000** | **1200** | **0.72** | **62%** |

capacitance $C_0$ and minimize switching losses [11]. Maximum power output occurs near series resonance, and as the operating frequency increases, output power decreases and converter efficiency increases [10]. In essence, piezoelectric power conversion utilizes the piezoelectric resonator as the converter's sole energy storage element [Fig. 1 (b)].

Although the working principle has been proven [12]–[15], the piezoelectric power converter's performance is limited by the integrated resonator, specifically by moderate quality factor ($Q$), electromechanical coupling ($k_t^2$), and spurious modes near resonance. Lower $Q•k_t^2$ reduces converter efficiency, while spurious modes between series and parallel resonances limit the converter's operating range [13].

However, conventional spurious suppression methods (e.g., apodization and raised/recessed frame [1], [16]–[18]) are insufficient, as they tend to spread out spurious modes or prove to be difficult to implement at MHz frequencies. Thus, we propose a novel spurious-free bulk acoustic resonator design in lithium niobate (LiNbO$_3$) that surpasses the state-of-the-art (SoA, Table 1) in spurious suppression with a high figure-of-merit (FoM, $Q•k_t^2$). Future research could extend this design to other applications, including filters, oscillators, transformers, etc.

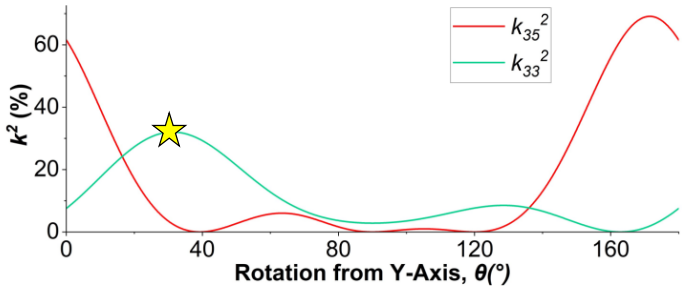

Fig. 2. Coupling coefficient, $k^2$, as the rotation from the y-axis varies, for TS mode ($k_{35}^2$) and TE mode ($k_{35}^2$). 36Y-cut LiNbO$_3$ was selected for this work, marked by the star.

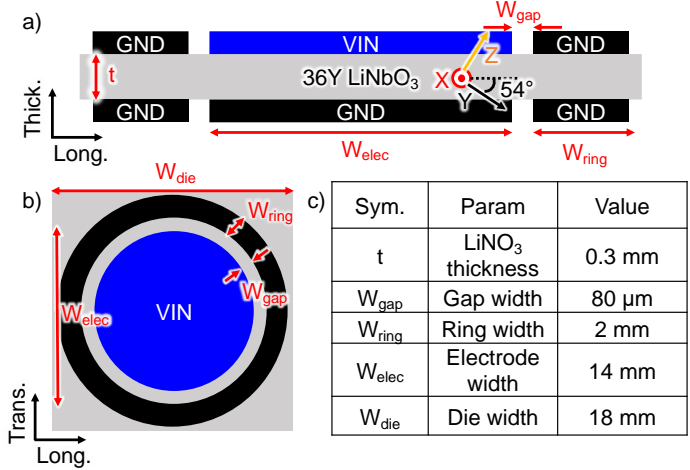

Fig. 3. Illustration of (a) side-view and (b) top-view of the proposed resonator design with grounded ring, with parameters tabulated in (c). All electrodes have aluminum (Al) thickness of 300 nm.

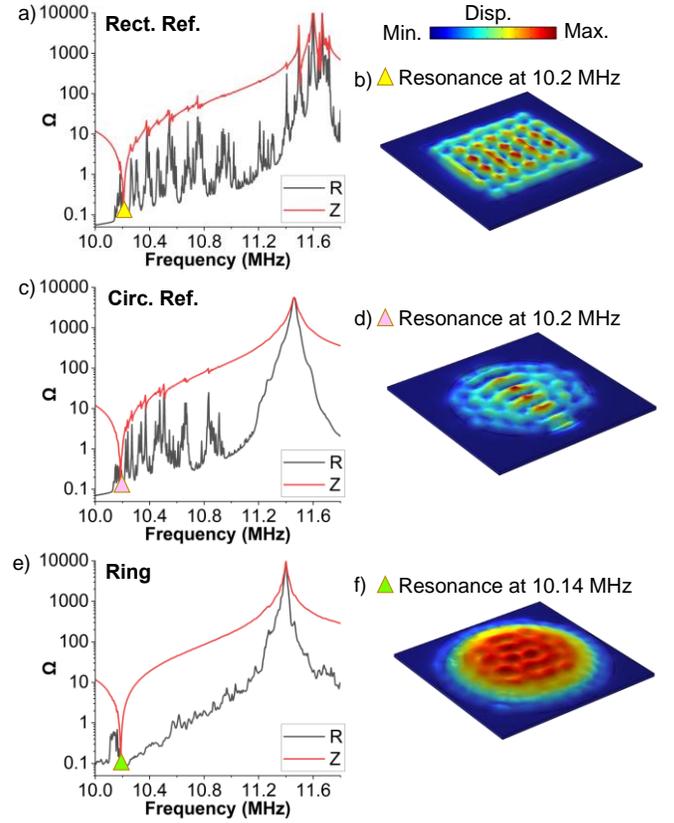

Fig. 4. Simulated impedances (Z) and resistances (R) of the rectangular reference design (a), the circular reference design (c), and the novel grounded ring design (e). Displacement at resonance of all three designs, marked by triangles in the impedance plots, are illustrated in (b, d, f).

## II. DESIGN & SIMULATION

The bulk acoustic, thickness-extensional (TE) resonator was designed using LiNbO$_3$ for its intrinsic high electromechanical coupling and low loss material properties [19]–[24]. However, selecting the orientation and direction of the applied electric field is challenging as LiNbO$_3$ is highly anisotropic [25]. The goal is to choose a crystal orientation that increases coupling of the TE mode to induce uniform vibration, thus increasing Q and minimizing parasitic couplings. Fig. 2 plots $k^2$ as a function of the rotated Y-axis, namely $k_{33}^2$ ($e_{33}^2/c_{33}/\varepsilon_{33}$) for TE and $k_{35}^2$ ($e_{35}^2/c_{55}/\varepsilon_{33}$) for thickness shear (TS). In order to optimize TE coupling while minimizing other modes, 36Y-cut LiNbO$_3$ was selected as a commercially viable option. Thanks to its unique dispersion behavior, 36Y-cut LiNbO$_3$ is an optimal choice for confining energy of the TE mode [26]–[28].

A novel "grounded ring" structure is proposed for spurious suppression. The novel resonator features center electrodes on the top and bottom that are electrically excited in opposing configurations. These electrodes are further surrounded by a non-metallized separation gap, which are then surrounded by a metallized "ring" that is electrically grounded on top and bottom [Fig. 3 (a)]. From the top-view, the center electrode, non-metallized separation gap, and ring are circularly-shaped for spurious suppression [Fig. 3 (b)]. Dimensions are enumerated in Fig. 3 (c).

The dimensions of the grounded ring and separation gap were optimized via parametric sweep. It was found that a smaller separation gap generally improved performance but posed potential challenges with power handling, while a larger ring width improved performance yet saturated after a certain threshold was reached.

In principle, the separation gap generated by the grounded ring not only maintains the TE mode, but also eliminates lateral spurious tones. Unlike [29], where a recessed frame in a certain structure removes lateral modes by altering the dispersion characteristics, careful implementation of our design uses a grounded ring for spurious-free operation by electrically loading the piezoelectric material such that it can reinforce the TE coupling [30]. In conjunction, the circular shape leverages the isotropic piezoelectric coefficient $e_{33}$ while suppressing the anisotropic $e_{31}$ in 36Y-cut LiNbO$_3$ [31].

This novel design was simulated using three-dimensional (3D) finite element analysis (FEA) in COMSOL. For comparison, two reference designs were also simulated. First, a rectangular reference TE design is shown, consisting of rectangular electrodes centered on the top and bottom of LiNbO$_3$. The simulated impedance and resistance reveal large spurious modes in the inductive region of the resonator, making it virtually unusable for piezoelectric power conversion [Fig. 4 (a)]. The displacement mode shape reveals multiple modes at resonance, indicating non-uniform vibration [Fig. 4 (b)].

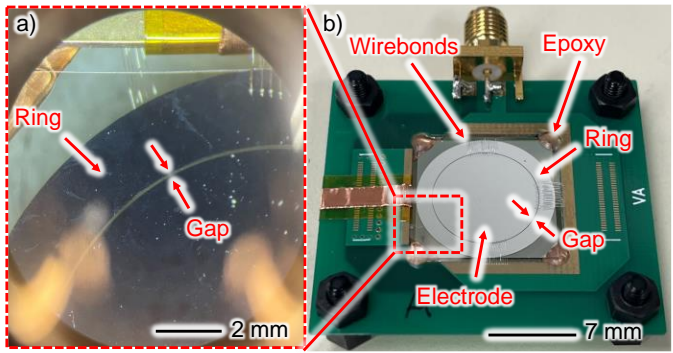

Fig. 5. Zoomed-in view (a) of the fabricated device epoxied and wire bonded to the substrate (b).

Second, a circular reference TE design is shown, where the electrodes are designed to be circular. While the active area has decreased compared to the rectangular reference design, resulting in a slight increase in resistance $R_m$, there is some spurious suppression near the parallel resonance [Fig. 4 (c)]. While the circular electrode shape mitigates some spurious modes by leveraging the anisotropy in $LiNbO_3$ [Fig. 4 (d)], the resonator still needs to further suppress lateral wave propagation, especially near resonance.

Lastly, our proposed design, which improves upon the circular TE design by adding the grounded ring, is simulated. The impedance and resistance are completely spurious-free [Fig. 4 (e)], thus increasing the spurious-suppressed region for increased converter operation range. The ring design vibrates with much more uniformity and greater amplitude, with little-to-no lateral mode shapes detected [Fig. 4 (f)].

### III. FABRICATION

After the design is thoroughly validated in FEA, the proposed resonator is fabricated with standard cleanroom procedures. Lithography is performed on 4-inch 0.3 mm thick 36Y-cut $LiNbO_3$, provided by Precision Micro-Optics, to form the electrode and ring patterns [32]. Afterwards, 300 nm of aluminum (Al) is deposited on both sides with an e-beam evaporator. The wafer thickness was chosen based on the frequency specifications set by the desired power converter operation. A clearly defined non-metallized gap separates the electrode from the ring [Fig. 5 (a)]. The wafer is then diced and the individual resonator is epoxied at the corners and wire bonded to the testbed [Fig. 5 (b)]. The resonator itself is 18 x 18 $mm^2$ in size, while the entire mounted device has an area of 28 x 28 $mm^2$. Copper traces are routed to an SMA connector for characterization.

### IV. RESULTS

The measured impedance, resistance, and Bode $Q$ [33] of the rectangular reference and novel designs are compared in Fig. 6. The rectangular reference design [Fig. 6 (a)] features the same design as that in Fig. 4 (a). The measured results show huge spurious modes that are greatly suppressed in the ring design [Fig. 6 (b)]. The remaining spurious modes in the proposed design are likely caused by wafer thickness variations. Both fabricated devices demonstrate Bode $Q$ around 4000 [Fig. 6 (c-d)]. Since Bode Q depends on group delay, it is extremely sensitive to spurious modes. The reference design [Fig. 6 (c)] yields an inconsistent value of $Q$ as the frequency varies. In contrast, our proposed design [Fig. 6 (d)] measures a smoother and more constant $Q$ over a broader frequency range, suggesting a completely spurious-free performance.

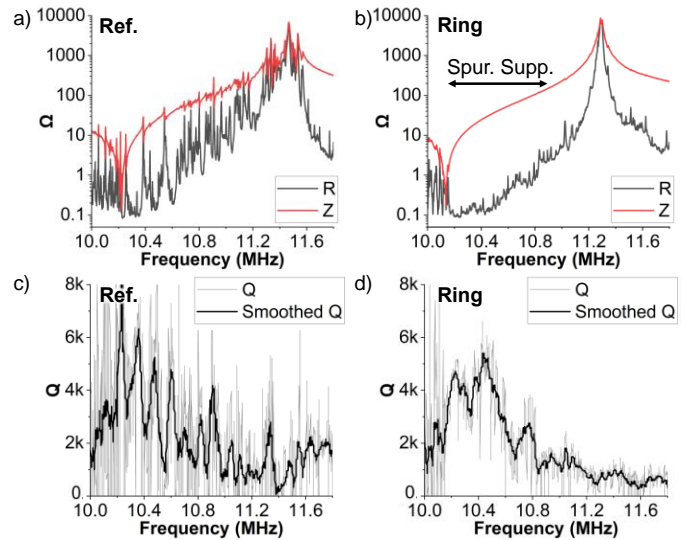

Fig. 6. Measured impedance/resistance and Bode Q for the reference device (a,c) and the proposed ring device (b,d), with the spurious-suppressed region highlighted in yellow.

Finally, the proposed design features $k_t^2$ of 30% with a spurious-suppressed region of 0.72 MHz and a fractional suppressed region of 62%. The spurious-suppressed region is defined as the frequency range where resistance is no larger than 20 x $R_m$ (minimum resistance), and the fractional suppressed region is the ratio of the spurious-suppressed region to the difference between series and parallel resonance frequencies. These metrics aim to characterize spurious suppression over a range of frequencies; wider spurious suppression expands the converter's output powers.

The proposed bulk acoustic resonator surpasses the SoA spurious suppression methods (Table 1) with the highest fractional suppressed region of 62%, while maintaining a high FoM of 1200. Thus, our design shows great potential for not only piezoelectric power conversion, but also any application requiring high FoM and no spurious modes.

### V. CONCLUSION

This work reports a spurious-free bulk $LiNbO_3$ acoustic resonator design for piezoelectric power conversion with high $Q$ of 4000, $k_t^2$ of 30%, and a large fractional suppressed region of 62%. First, the optimal $LiNbO_3$ orientation cut was selected based on its ability to maximize the TE mode. Then, a novel resonator topology was designed and extensively validated against existing conventional designs. Lastly, this design was fabricated and characterized, showing excellent results. Future research could extend the proposed grounded ring-based spurious-free resonator design to other applications, such as filters, oscillators, and transformers.